\documentclass[twocolumn,final,aps,prl,showpacs,superscriptaddress,amsmath,amssymb,amsfonts,floatfix]{revtex4-1}

\usepackage{graphicx}
\usepackage{multirow}
\usepackage{epsfig}
\usepackage{url}
\usepackage{color}
\usepackage{grffile}

\usepackage[colorlinks,breaklinks,bookmarks=true,citecolor=blue,linkcolor=blue,urlcolor=blue]{hyperref}
\usepackage{color}
\usepackage{tabularx}
\usepackage{sidecap}
\usepackage{soul}
\usepackage{float}

\begin{document}

\title{Impact of rotational symmetry breaking on \texorpdfstring{$d$}--wave superconductivity in Hubbard models for cuprate and nickelate superconductors}

\author{Hongdao Zhuge}
\affiliation{School of Physical Science and Technology, Soochow University, Suzhou 215006, China}

\author{Liang Si}
\email{siliang@nwu.edu.cn}
\affiliation{School of Physics, Northwest University, Xi'an 710127, China}
\affiliation{Institute of Solid State Physics, TU Wien, 1040 Vienna, Austria}

\author{Mi Jiang}
\email{jiangmi@suda.edu.cn}
\affiliation{School of Physical Science and Technology, Soochow University, Suzhou 215006, China}
\affiliation{Jiangsu Key Laboratory of Frontier Material Physics and Devices, Soochow University, Suzhou 215006, China}

\begin{abstract}
Recent experiments have revealed the substantial impact of broken rotational symmetry on the superconductivity. In the pursuit of understanding the role played by this symmetry breaking particularly in cuprate and nickelate superconductors on their superconductivity, we investigated two characteristic symmetry breaking mechanisms arising from (1) structurally orthogonal distortions from $C_4$ to $C_2$ symmetry and (2) anisotropic hybridization between $d_{x^2-y^2}$ orbital and an additional metallic band within the framework of the Hubbard model by employing dynamic cluster quantum Monte Carlo calculations. We discovered that the anisotropy is generically detrimental to the $d$-wave pairing so that the experimental findings of much lower superconducting $T_c$ of infinite-layer nickelates compared with the cuprates may be connected to the intrinsic anisotropy. Our exploration sheds light on the fundamental anisotropy factors governing superconductivity in nickelates and cuprates and offer insights contributing to the broader understanding of unconventional superconductors in anisotropic environment.
\end{abstract}

\maketitle

{\em \textcolor{blue}{Introduction:}}
Unconventional superconductivity (SC) in cuprate~\cite{keimer2015} and nickelate~\cite{2019Nature,Pr,La,LaSC,Nd6Ni5O12} superconductors are widely believed to mainly originate from the electron-electron interactions that have been described by the Hubbard models~\cite{Karp2020,Kitatani2020,Held2022,kitatani2022optimizing}. 
The smoking-gun evidence of the pairing mechanism in both nickelate and cuprate superconductors are still lacking, albeit the recent experiments \cite{cheng2024evidence,sun2024electronic,ding2024cuprate} and theoretical studies \cite{Nomura2019,Kitatani2020} indicate that the single Ni-$d_{x^2-y^2}$ band in nickelates appears to perform a role akin to that in cuprates despite that the introduction of an extra $s$-band \cite{jiang2022characterizing} and the consequent inter-orbital hybridization could intricately impact the electronic structure and thereby alter the $T_c$ \cite{jiang2022PhysRevB.106.224517}.
The electronic structure calculations \cite{gu2020substantial} demonstrated that the rather broad interstitial $s$-band crossing the Fermi energy (E$_f$) is composed of a combination of orbitals including Nd-5$d_{xy}$ and 4$f$, Ni-$d_{yz}$/$d_{xz}$, O-2$p$ and other interstitial states. Intriguingly, this broad band results in the formation of small electron pockets at $A$ point of Brillouin zone, which suggests that the infinite-layer nickelates have distinct low-energy physics from the cuprates when describing, e.g., Hall conductivity \cite{Aritareview, ChenHMillis2022Front}.

A possible consequence induced by the hybridization between $d_{x^2-y^2}$ and other conducting bands is the absent $C_4$ symmetry \cite{gu2020substantial,wang2022rare,ji2023rotational,PhysRevB.106.035111}, if the doped cations do not host such a conducting band, e.g., 5$f$ orbitals/bands are absent in Sr (Ca) in (Nd,Sr)NiO$_2$ [(La,Ca)NiO$_2$] while present in Nd (La). Consequently, the non-zero hopping/hybridization induces not only a self-doping effect and Kondo effect \cite{Zhang2019}, but also a possible symmetric reduction via anisotropic distribution of doped cations and hybridization  \cite{gu2020substantial,ChenHMillis2022Front}. As to the roles played by the surrounding bands near E$_f$,
in the last decades, some proposals have suggested that the ``incipient'' bands, namely full (empty) bands slightly below (above) the Fermi energy, can significantly enhance $T_c$ owing to either inter-band pair-scattering channels \cite{PhysRevB.72.212509, PhysRevResearch.4.013032} or contributing to the spin fluctuation \cite{PhysRevLett.117.077003}; while other study pointed out that the incipient band is not beneficial for enhancing $T_c$ \cite{PhysRevB.104.245109}.

The second alternative driver of symmetry breaking in cuprate and nickelate superconductors could be the distortion of the real-space crystal lattice. The commonly utilized Emery or Hubbard models are typically formulated on the premise of a $C_4$ symmetry lattice characterized by the isotropic band dispersion. However, experimental observations in cuprates and nickelates, such as the manifestation of charge order \cite{PhysRevLett.129.027002,tam2021charge,PhysRevX.13.011021}, stripe phase \cite{PhysRevLett.122.247201} or anisotropy in critical fields \cite{chow2023dimensionality,wang2022rare,PhysRevB.107.L220503}, point to the potential absence of the rotational $C_4$ symmetry.

Hence, the key facet in unraveling the enigma of high-temperature nickelate and cuprate SC lies in understanding the intricate interplay between electronic structure and symmetry breaking and their further impact on the critical temperature. Symmetry breaking, whether arising from structural distortions or electronic interactions and/or hybridization, induces a profound influence on the electronic properties, and even extends to the pairing mechanism responsible for the emergence of SC. However, a comprehensive exploration of the effects of gradual symmetry breaking on critical temperature is nearly unfeasible through experiments. This is due to the practical challenge of achieving a continuous and gradual change in the degree of symmetry breaking during material synthesis.
In this context, exploring the impacts of symmetry breaking on electronic structure and SC within the framework of theoretical models like the celebrated Hubbard model becomes feasible and demanded. Such investigation can not only deepen our understanding of the fundamental physics governing these materials but also hold the potential to unlock new frontiers in the design and development of novel superconducting materials.

The above discussions raise the following questions: (1) how many kinds of symmetry breaking are possible to be simulated in Hubbard model? (2) How to describe different type of symmetry breaking in Hubbard model accurately? Answering these questions is the prerequisite to demonstrate how the rotational symmetry breaking quantitatively affect the superconducting properties.
To answer these questions, we explored two characteristic models with 2D Hubbard interactions representing the possible rotational symmetry breaking in superconducting cuprates and nickelates: (1) anisotropic nearest-neighbor hopping integrals and (2) anisotropic hybridization (hopping) between $d_{x^2-y^2}$ to a conducting (interstitial) $s$-band. Both models characterize the symmetry reduction from $C_4$ to $C_2$, and our numerical simulations employing the dynamic cluster quantum Monte Carlo calculations demonstrate that both of them have significant impact on the SC.

{\em \textcolor{blue}{Model and Method:}}
As the first model, we employ the two-dimensional (2D) Hubbard model with anisotropic hopping integrals along $\hat{x}$ and $\hat{y}$ directions on the square lattice
\begin{align} \label{eq:HM}
	\hat{H}_{hub} = & \sum_{k\sigma} E_k n^d_{k\sigma}  + U \sum_i n_{i\uparrow}n_{i\downarrow}
\end{align}
where the non-interacting dispersion $E_k=-2 (t_x \cos k_x+ t_y \cos k_y)+4 t'_d \cos k_x\cos k_y -\mu$ incorporates the generic anisotropic hoppings with $t_x \neq t_y$ and conventionally isotropic case $t_x = t_y$ as well. 
Our goal is to systematically investigate the impact of the ratio $t_y/t_x$ (with fixed $t_x=1$) on the $d$-wave pairing instability. 
Note that the anisotropic Hubbard model has been studied in the context of the ladder-type lattices~\cite{YANG2022128316}.

In addition, we investigated the two-orbital $d$-$s$ model with correlated $d$ orbital and uncorrelated metallic $s$ orbital coupled with each other~\cite{jiang2022characterizing} 
\begin{align} \label{eq:HM}
	\hat{H}_{ds} = & \sum_{k\sigma} (E^d_k n^d_{k\sigma} +E^s_k n^s_{k\sigma}) + U \sum_i n^d_{i\uparrow}n^d_{i\downarrow}\nonumber\\ 
	& + \sum_{k\sigma} V_k (d^\dagger_{k\sigma}s^{\phantom\dagger}_{k\sigma}+h.c.) 
\end{align}
with two orbitals' dispersion and $d$-$s$ hybridization as
\begin{align} \label{dis}
	E^d_k = & -2 t_d (\cos k_x+\cos k_y)+4 t'_d \cos k_x\cos k_y -\mu \nonumber\\ 
	E^s_k = & -2 t_s (\cos k_x+\cos k_y) + \epsilon_s -\mu \nonumber\\ 
	V_k = & -2 (V_x \cos k_x+ V_y \cos k_y)
\end{align}
where $d^{\dagger}_{k\sigma}(s^{\dagger}_{k\sigma})$ are electronic creation operators in momentum space for two orbitals. $n^{d(s)}_{i\sigma}$ and $n^{d(s)}_{k\sigma}$ are the associated number operators in the real and momentum spaces separately. The chemical potential $\mu$ tunes the total electron density while $\epsilon_s$ controls the relative density between two orbitals. 
We will explore the distinct behavior between anisotropic ($V_x \neq V_y$) and isotropic ($V_x= V_y$) $d$-$s$ hybridizations.
Note that we focus on the situation of dilute limit $n_s \sim 0.1$ as in our previous work~\cite{jiang2022PhysRevB.106.224517}, where the $d$-wave superconducting dome was uncovered to shift to the overdoped regime.
Its relevance to the infinite-layer nickelate superconductors is manifested by the recent DFT calculation with Wannier fitting, which indicated that the dominant hybridization between Ni-3$d$ orbitals and the effective interstitial s orbital can induce a large inter-cell hopping as mimicked by our hybridization $V_k$~\cite{ChenHGY2020}. Moreover, the high-pressure enhancement~\cite{wang2022pressure} of $T_c$ has direct implication to our $d$-$s$ model with $V_{x/y}$ mimicking the pressure effects. 
DFT calculations revealed that the infinite-layer nickelates can be described by the extremely anisotropic case of $V_{x/y}\sim$0 \cite{Si2019,Kitatani2020}. Therefore, we will focus on this situation compared with the conventional isotropic hybridization. Without loss of generality, we adopt $V_y=0$ as the convention and tune $V_x$.

We adopt dynamical cluster approximation (DCA)\newline ~\cite{Hettler98,Maier05,code} with the continuous-time auxilary-field (CT-AUX) quantum Monte Carlo (QMC) cluster solver~\cite{GullCTAUX} to
numerically solve both models. 
As a celebrated quantum many-body numerical method, DCA calculates the physical quantities in the thermodynamic limit via mapping the bulk lattice problem onto a finite cluster embedded in a mean-field bath in a self-consistent manner~\cite{Hettler98,Maier05}. DCA and its cousins of quantum embedding methods have provided much insights on the strongly correlated electronic systems~\cite{Maier05}, despite of its limitations originating from the smaller tractable cluster size than finite-size QMC simulations albeit with better minus sign problem. 

The SC properties can be studied via solving the Bethe-Salpeter equation (BSE) in the eigen-equation form in the particle-particle channel~\cite{Maier2006,scalapino2007numerical}
\begin{align} \label{BSE}
    -\frac{T}{N_c}\sum_{K'}
	\Gamma^{pp}(K,K')
	\bar{\chi}_0^{pp}(K')\phi_\alpha(K') =\lambda_\alpha(T) \phi_\alpha(K)
\end{align}
where $\Gamma^{pp}(K,K')$ denotes the lattice irreducible particle-particle vertex of the effective cluster problem with combining the cluster momenta $\bf K$ and Matsubara frequencies $\omega_n=(2n+1)\pi T$ as $K=(\mathbf{K}, i\omega_n)$. 

The normal state pairing tendency is reflected by the leading eigenvalue $\lambda_\alpha(T)$ for pairing symmetry $\alpha$. Simultaneously, the associated eigenvector $\phi_\alpha(K)$ can be viewed as the normal state analog of the SC gap function~\cite{Maier2006,scalapino2007numerical}.
It has been widely accepted that the $d$-wave pairing plays a dominant role in the cuprate superconductors and closely relevant Hubbard model~\cite{scalapino2007numerical,Qin2022AnnualRev}. Therefore, instead of nematic and/or stripe ordering tendency that has been extensively investigated within Hubbard-type models, here we only focus on the $d$-wave pairing symmetry.


\begin{figure} 
\psfig{figure=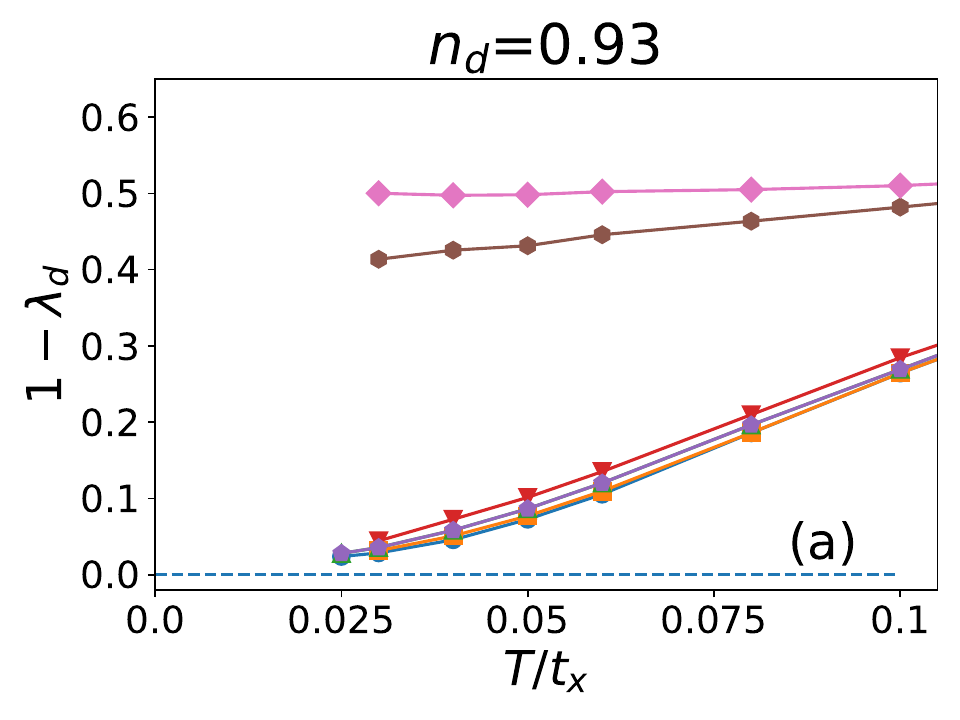,height=3.2cm,width=.23\textwidth, clip} 
\psfig{figure=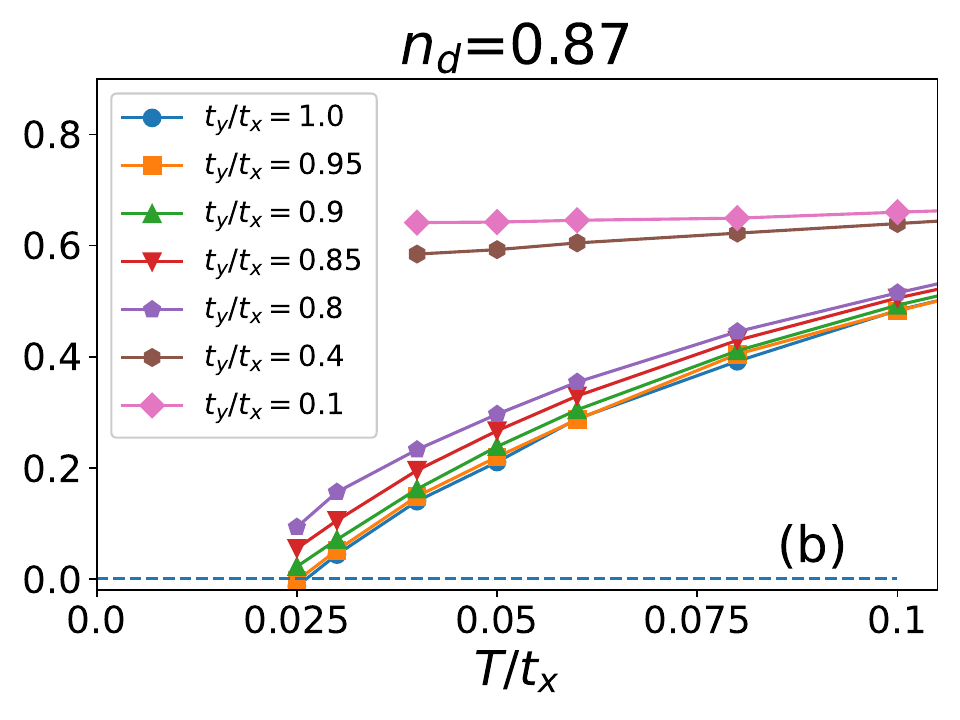,height=3.2cm,width=.22\textwidth, clip} 
\psfig{figure=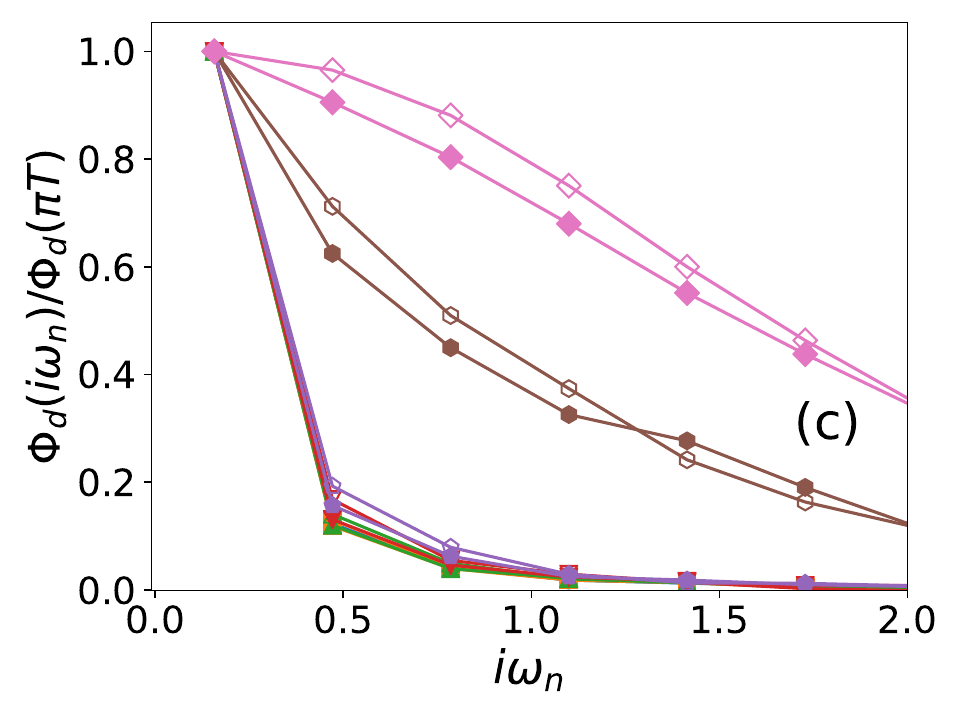,height=3.0cm,width=.23\textwidth,trim=12 0 13 0, clip} 
\psfig{figure=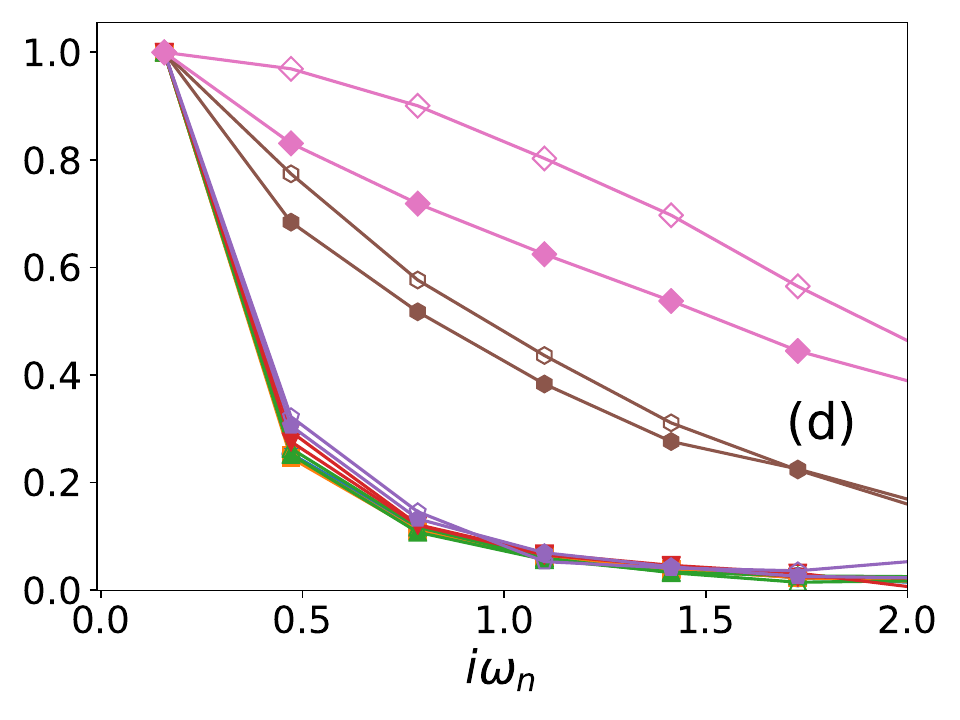,height=3.0cm,width=.22\textwidth,trim=12 0 13 0, clip} 
\caption{(Upper) leading $d$-wave pairing eigenvalue of BSE versus $T$ and (Lower) associated eigenvector (normalized by the value at $i\omega_n=\pi T$) for at ${\bf K}=(\pi,0)$ (filled) and ${\bf K}=(0,\pi)$ (unfilled) at $T/t_x=0.05$ of the single band anisotropic Hubbard model. The DCA cluster size is $N_c=8$.}
\label{ldhub}
\end{figure}

{\em \textcolor{blue}{Single-orbital model:}}
To investigate the influence of symmetry breaking on the SC properties, we first focus on the temperature evolution of the leading $d$-wave $\lambda_d(T)$ of the anisotropic Hubbard model in Figure~\ref{ldhub} (upper panels) for varied anisotropy $t_y/t_x$ at two characteristic fillings $\rho=$0.93 and 0.87.
Apparently, the anisotropy results in the gradual destruction of $d$-wave pairing tendency that is manifested by the departure of $1-\lambda_d(T)$ curves from zero, which generically matches with the trend found in previous calculations~\cite{YANG2022128316}. In particular, the impact of the anisotropy is not obvious until $t_y/t_x < 0.9$, which seemingly implies that sufficiently large anisotropy is required to induce significant effects on the $d$-wave pairing.

Although $T_c$ can be determined by $\lambda_d(T_c)=1$ in principle, it is practically challenging to simulate low temperatures close enough to $T_c$. Nonetheless, the monotonic decrease of $T_c$ with the magnitude of anisotropy is obvious for both fillings from $1-\lambda_d(T)$. Note that the presence of the anisotropy does not qualitatively modify the  distinct temperature variations between $\rho=$0.93 and 0.87. Precisely, at higher density $\rho=$0.93 associated with the pseudogap features without the anisotropy, $1-\lambda_d(T)$ shows exponential behavior at lowest temperature regime; while at $\rho=$0.87, the logarimic dependence remains in the anisotropic situations~\cite{Maier19}. 

The lower panels of Fig.~\ref{ldhub} further illustrate the frequency dependence of the leading eigenvector $\phi_d$ at ${\bf K}=(\pi,0)$ (filled) and ${\bf K}=(0,\pi)$ (unfilled) whose degeneracy in the isotropic model has been destroyed, where the negligible effects at small anisotropy $t_y/t_x > 0.9$ mirrors the features of $1-\lambda_d(T)$ discussed above.
The retardation nature of the $d$-wave pairing interaction is reflected by the decaying of $\phi_d$ with a characteristic frequency scale. Firstly, the decaying rate is slower by turning on a large enough anisotropy $t_y/t_x > 0.9$ for both antinodal ${\bf K}$ directions and both fillings. Secondly, the comparison between $\Phi_d$ and $1-\lambda_d(T)$ at $T/t_x=0.05$ reveals that the generally slower decay at $\rho=$0.87 has weaker pairing tendency. These two aspects again implies for the significance of the moderate retardation for optimal SC~\cite{jiang2022PhysRevB.106.224517}.

\begin{figure} 
\psfig{figure=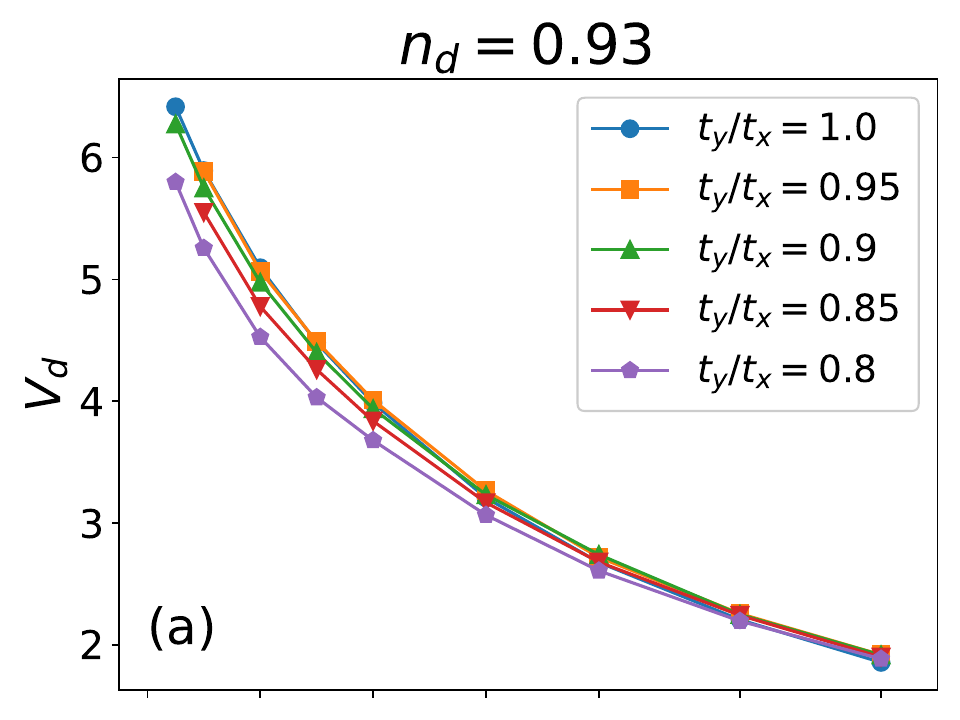,height=3cm,width=.23\textwidth,trim=-15 0 2 0, clip} 
\psfig{figure=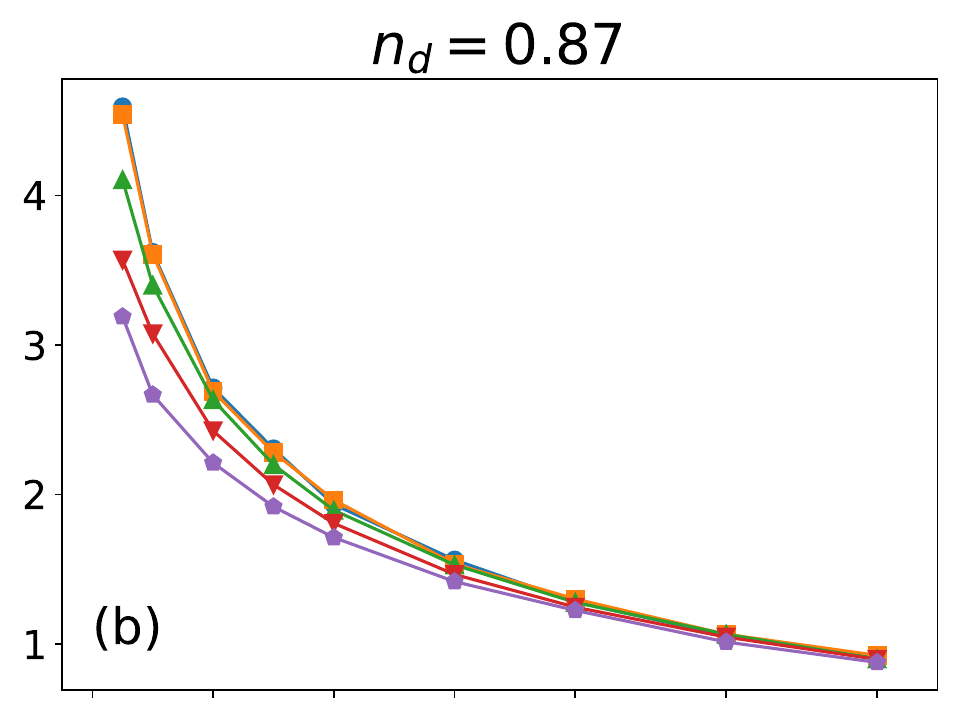,height=3cm,width=.22\textwidth,trim=-15 0 2 0, clip} 
\psfig{figure=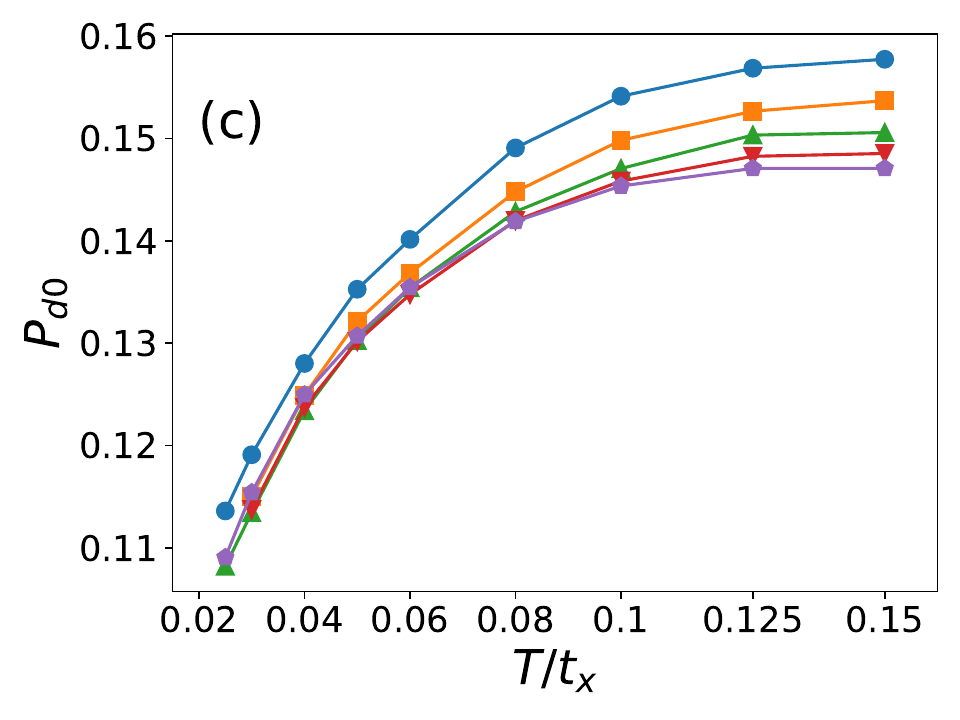,height=3cm,width=.23\textwidth,trim=13 0 2 5, clip} 
\psfig{figure=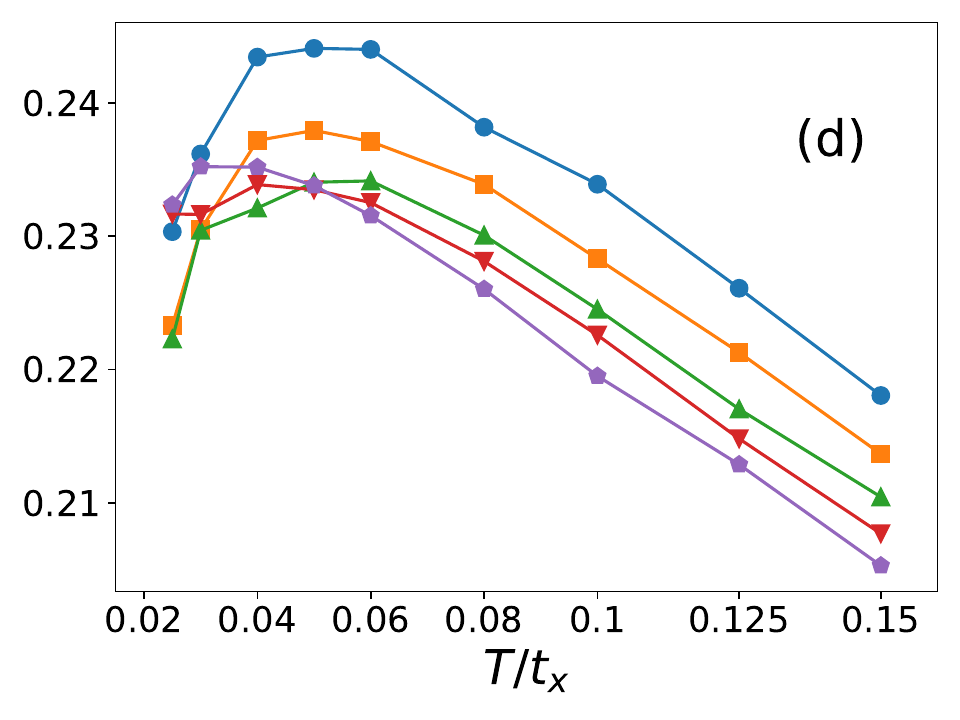,height=3cm,width=.22\textwidth,trim=13 0 2 0, clip} 
\caption{(Left) Temperature dependence of (Upper) effective $d$-wave pairing interaction $V_d$ and (Lower) bare pair-field susceptibility $P_{d0}$ for the single band anisotropic model.}
\label{VdPd0hb}
\end{figure}

To have a better understanding of the pairing tendency and its variation with the anisotropy, we resort to the $d$-wave projected effective pairing interaction $V_d(T) = P^{-1}_{d0}(T)-P^{-1}_{d}(T)$, where the pair-field susceptibility $P_{d}$ and its bare counterpart $P_{d0}$ can be obtained via the dressed and bare two-particle Green's functions~\cite{jiang2022PhysRevB.106.224517,Jarrell2001,PhysRevB.88.245110}. 
Essentially, $V_d$ and $P_{d0}$ are two decomposed factors with similar roles as $\Gamma^{pp}$ and $\bar{\chi}_0^{pp}$ in Eq.~\ref{BSE} respectively.
Fig.~\ref{VdPd0hb} illustrates the temperature evolution of $V_d$ and $P_{d0}$, which both decrease with turning on anisotropy so that the pairing tendency is naturally weakened. Note that even in the absence of anisotropy, $P_{d0}$ decreases with lowering temperature while $V_d$ blows up, which indicates that the pairing interaction plays the decisive role for $d$-wave pairing. Besides, the impact of anisotropy is more obvious for weakening $V_d$ at lower filling $\rho=$0.87, which coincides with the generally weaker pairing trend with higher hole doping and also matches with the behavior of $1-\lambda_d(T)$ in Fig.~\ref{ldhub}.

\begin{figure} 
\psfig{figure=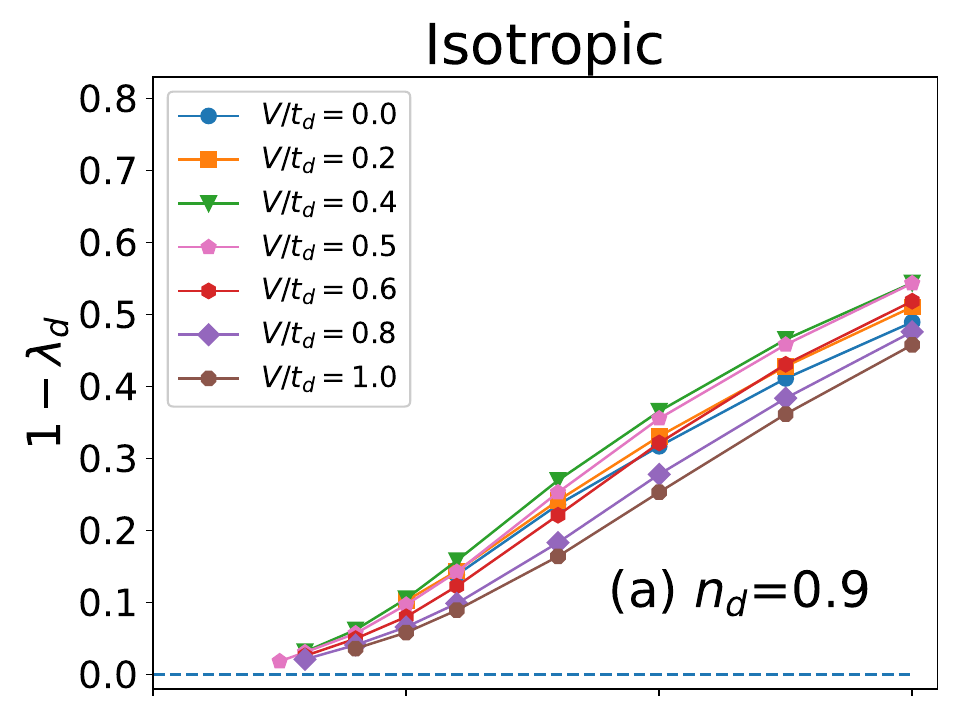,height=3cm,width=.24\textwidth,trim=5 2 4 0, clip} 
\psfig{figure=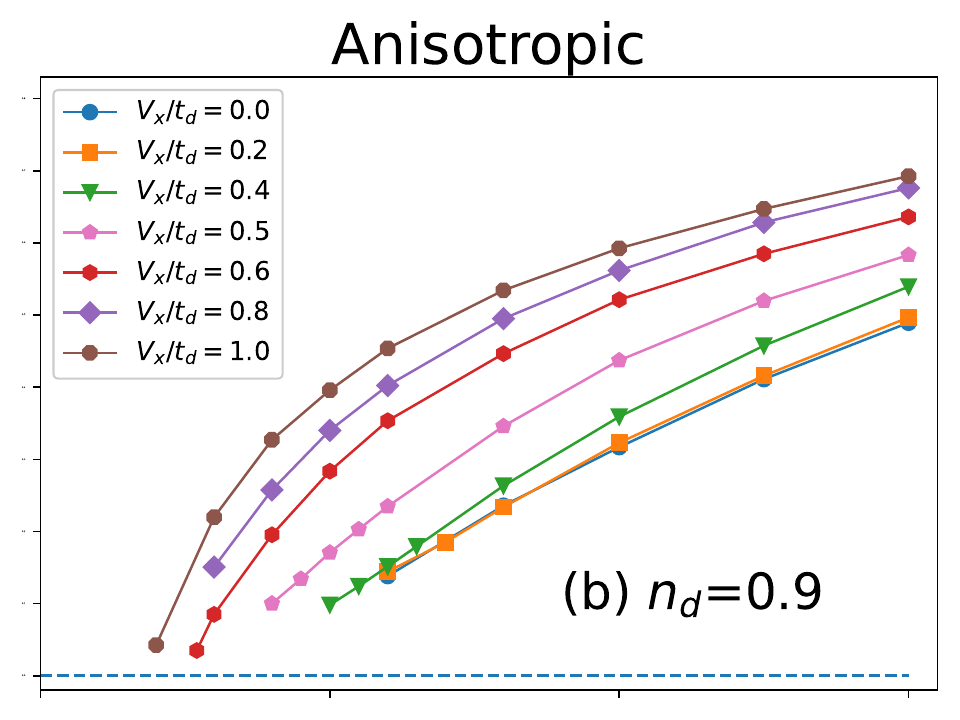,height=3cm,width=.22\textwidth,trim=15 0 -2 0, clip}  
\psfig{figure=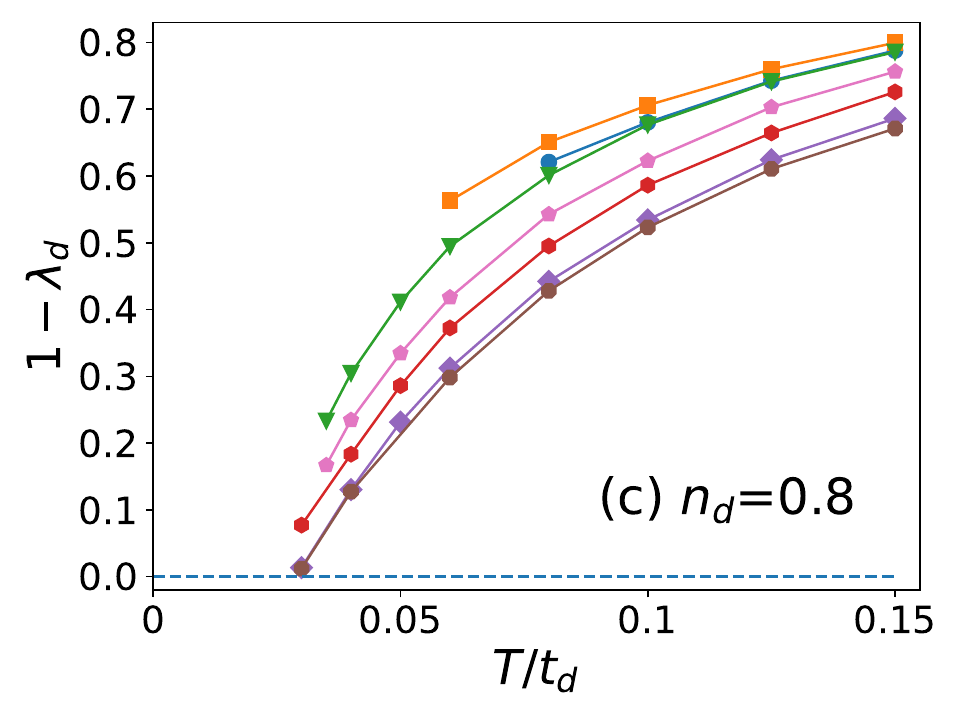,height=3.2cm,width=.24\textwidth,trim=7 -3 13 7, clip} 
\psfig{figure=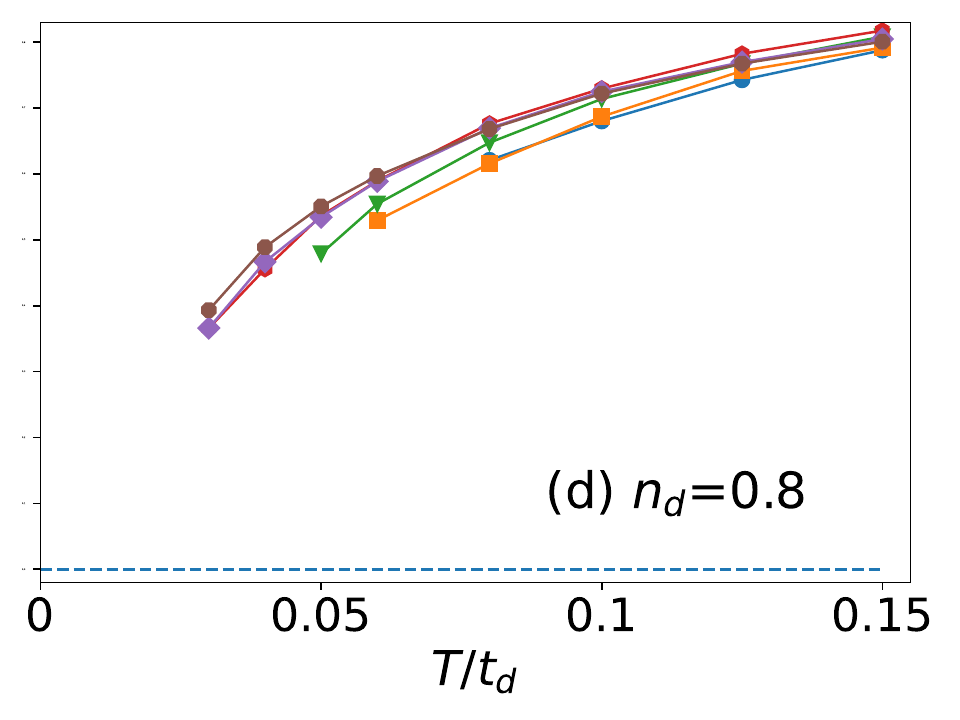,height=3.2cm,width=.22\textwidth,trim=15 0 11 7, clip}

\caption{Temperature evolution of $1-\lambda_d(T)$ for varied hybridization $V(V_x)$ at two characteristic fillings $n_d=$0.9 and 0.8 of $d$-$s$ model with fixed $n_s$=0.1 and $N_c=12$.}
\label{ldds}
\end{figure}

\begin{figure} 
\psfig{figure=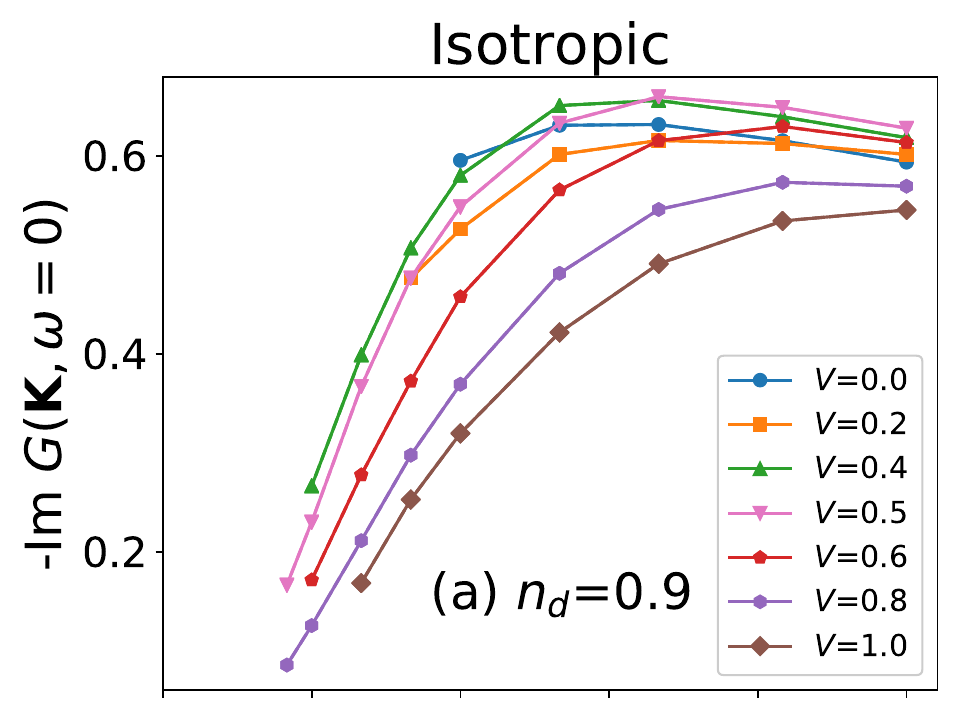,height=3cm,width=.24\textwidth,trim=0 0 5 0, clip} 
\psfig{figure=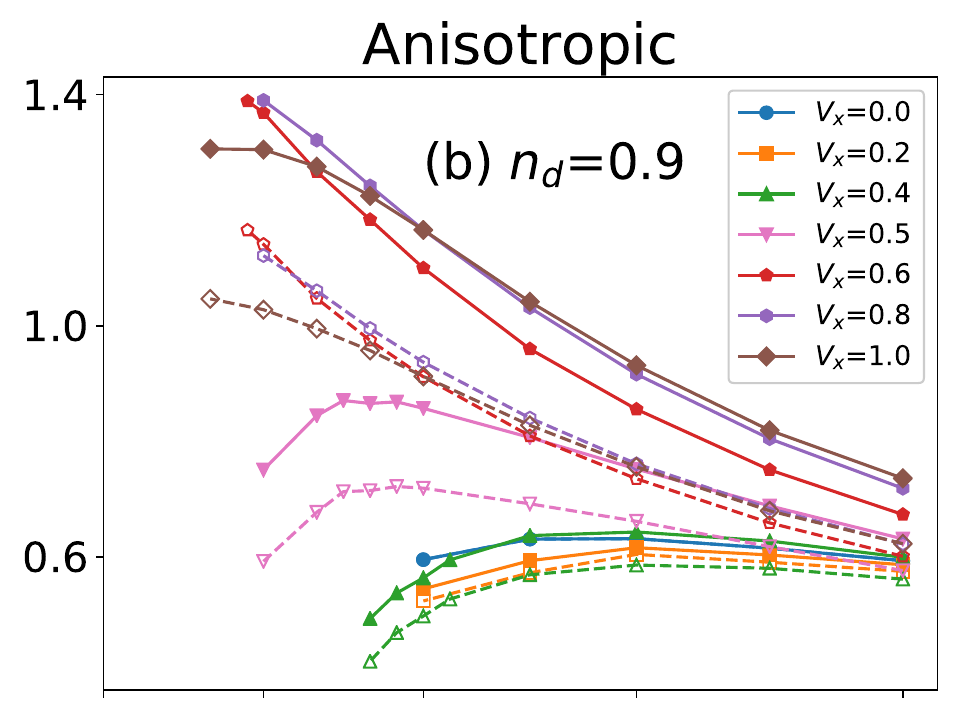,height=3cm,width=.22\textwidth,trim=8 0 4 0, clip} 
\psfig{figure=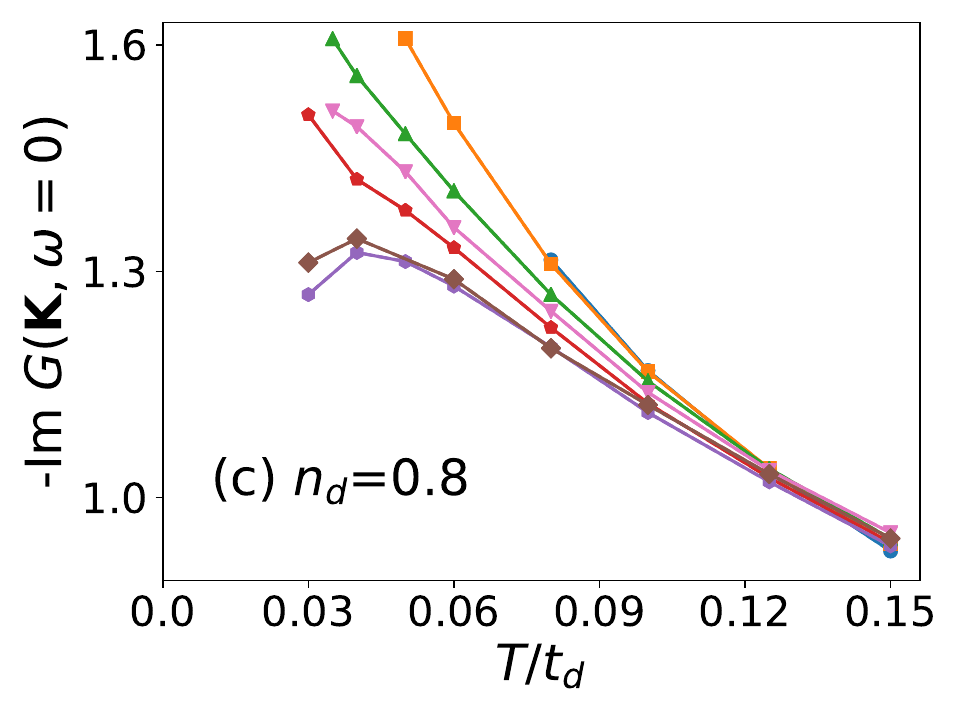,height=3.08cm,width=.24\textwidth,trim=1 8 14 10, clip} 
\psfig{figure=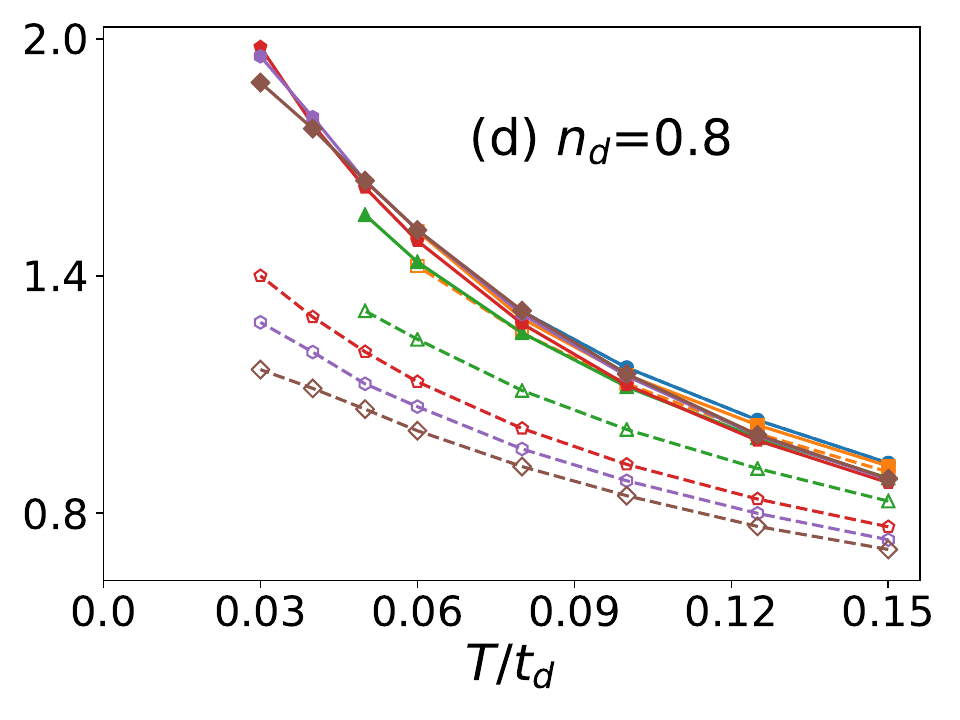,height=3.08cm,width=.22\textwidth,trim=8 8 13 12, clip} 

\caption{Extrapolated antinodal zero-frequency -ImG$(\bf{K}, \omega=0)$ obtained from a linear extrapolation of the first two Matsubara frequencies at $\bf{K}=(\pi,0)$ (solid) and $(0,\pi)$ (dashed).}
\label{Gkw}
\end{figure}

{\em \textcolor{blue}{Two-orbital model:}}
Now we switch to the two-orbital $d$-$s$ model. Our previous study revealed unusual shift of the SC dome with inter-orbital hybridization~\cite{jiang2022PhysRevB.106.224517}. Here we focus on the distinction between isotropic and anisotropic $d$-$s$ hybridization.

Figure~\ref{ldds} demonstrates the temperature evolution of $1-\lambda_d(T)$ for varied hybridization at two characteristic fillings $n_d=$0.9 and 0.8. The common difference between isotropic and anisotropic  hybridization, regardless of $n_d$, lies in that the former generically promotes while the latter suppresses the $d$-wave pairing within the accessible temperature regime above $T_c$. 

At higher $n_d=0.9$ (upper panels), in spite of the non-monotonic dependence on $V$ at relatively high $T/t_d\sim0.1$, the dominant feature of turning on isotropic $V$ is the exponential curvature change, which would suppress the extrapolated $T_c$ as a Berezinskii–Kosterlitz–Thouless (BKT) transition in two-dimensional system.
As discussed earlier~\cite{jiang2022PhysRevB.106.224517}, this indicates strong Emery-Kivelson phase. fluctuations~\cite{Maier2019,Emery1995} in the presence of pseudogap (PG). To illustrate the close relation between $1-\lambda_d(T)$ and the existence of PG, Figure~\ref{Gkw} (upper panels) provide evidence that the presence of a peak (whose temperature scale is normally assigned as $T^*$ of PG) of extrapolated antinodal zero-frequency -ImG$(\bf{K}, \omega=0)$ at $\bf{K}=(\pi,0)$ (solid) and $(0,\pi)$ (dashed) shows that the larger isotropic hybridization $V$ results in higher $T^*$, which implies for a possible higher $T_c$ in Fig.~\ref{ldds}(a). 

Nonetheless, in Fig.~\ref{ldds}(b), the anisotropy $V_x$ obviously induce the gradually stronger logarithmic behavior, which is supported by the disappearance of PG feature at sufficiently large $V_x$ in Fig.~\ref{Gkw}(b). Apparently, the anisotropy leads to the deviation between two antinodal directions despite that the $T^*$ scales are exactly the same.
In addition, there exists an abrupt change of $1-\lambda_d(T)$ around $V_x/t_d\sim0.5$, indicating drastic effects of strong enough anisotropic hybridization.

At smaller $n_d=0.8$, namely without PG behavior in the pure Hubbard model $V(V_x)=0$, the anisotropy has minor effects on the pairing tendency as shown in Fig.~\ref{ldds}(d). In spite of the opposite trends of Fig.~\ref{ldds}(c-d), $1-\lambda_d(T)$ remains its logarithmic evolution~\cite{jiang2022characterizing,Maier19} in all cases. 
Similarly, Fig.~\ref{Gkw}(c-d) display the absence of PG in most situations except for large isotropic $V$, where the peak implies that the evolution of $1-\lambda_d(T)$ in Fig.~\ref{ldds}(c) can switch to the exponential-like feature at even larger $V$ for simulations adopting larger DCA cluster $N_c$.

\begin{figure} 
\psfig{figure=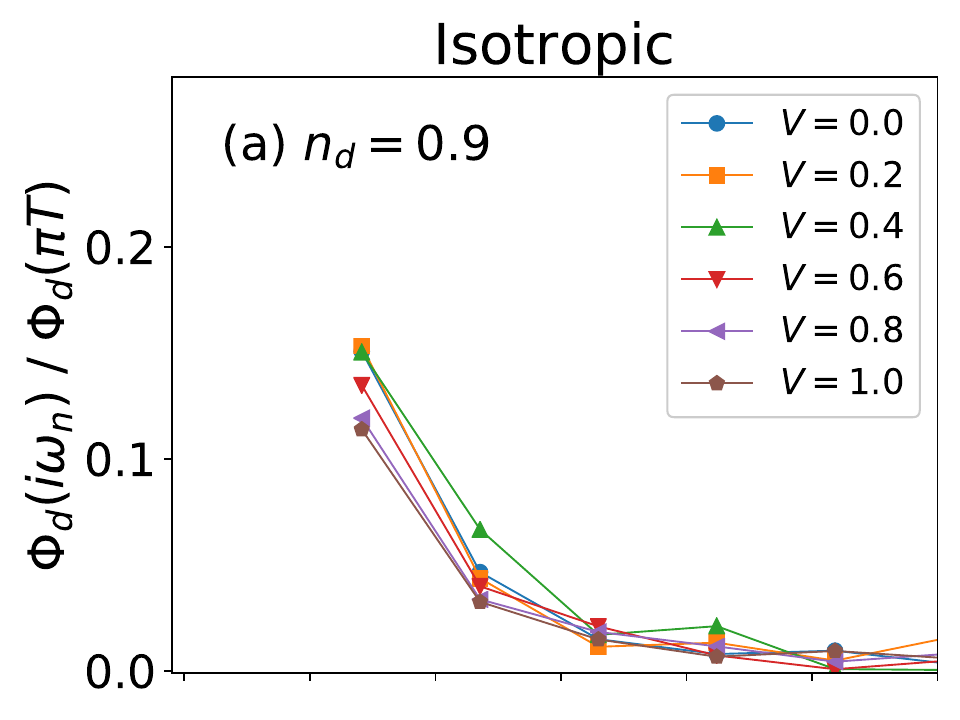,height=3.0cm,width=.24\textwidth,trim=3 9 5 0, clip} 
\psfig{figure=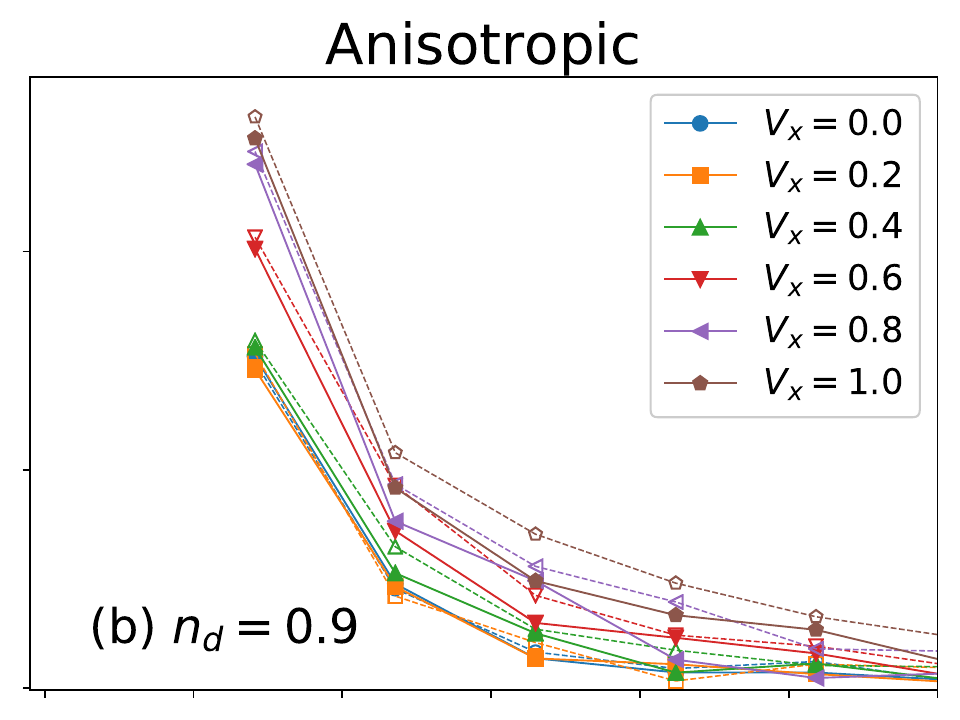,height=3.0cm,width=.22\textwidth,trim=10 1 2 -1, clip} 
\psfig{figure=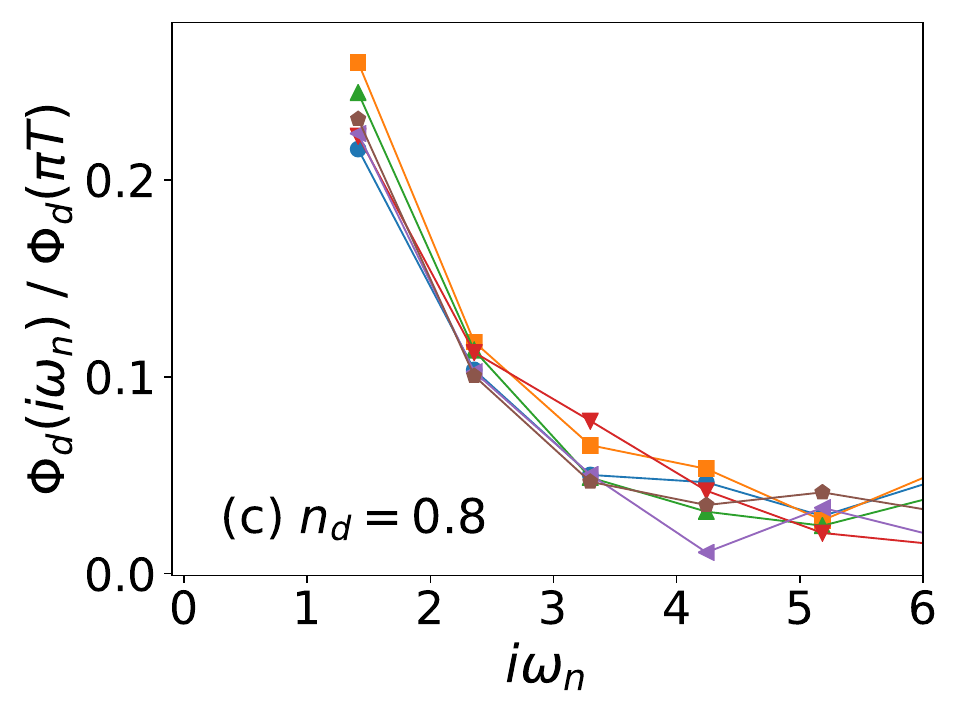,height=3.2cm,width=.24\textwidth,trim=4 0 12 9, clip} 
\psfig{figure=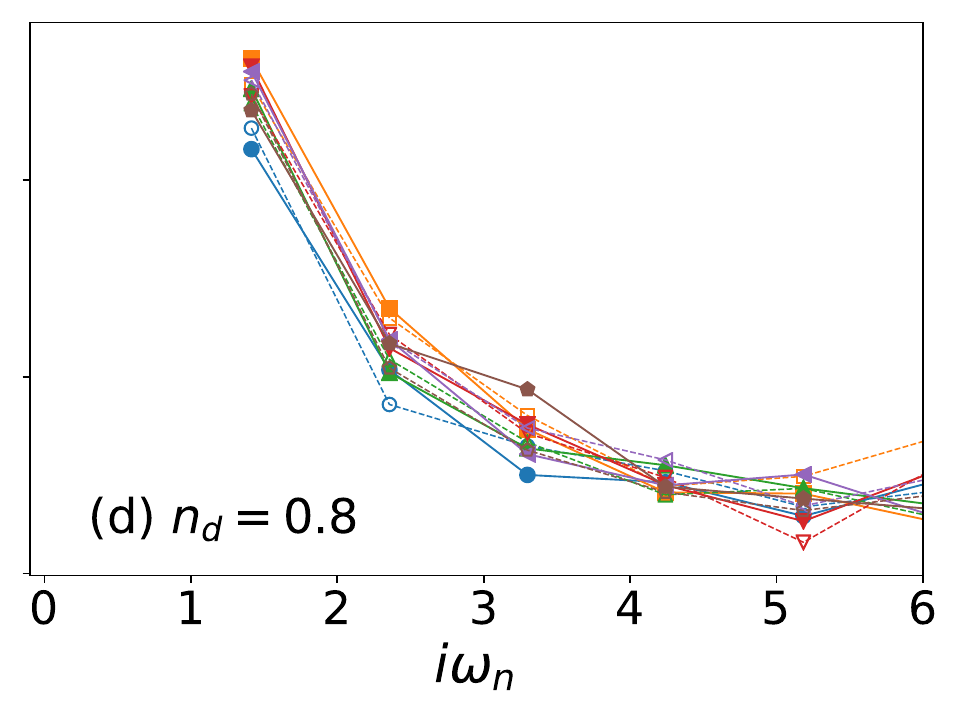,height=3.2cm,width=.22\textwidth,trim=10 0 10 9, clip} 
\caption{Leading $d$-wave eigenvector (normalized by the value at $i\omega_n=\pi T$ not shown) for ${\bf K}=(\pi,0)$ (filled) and ${\bf K}=(0,\pi)$ (unfilled) at $T/t_d=0.125$ of two-orbital $d$-$s$ model.}
\label{eighub}
\end{figure}

Akin to the single band model, Fig.~\ref{eighub} shows the frequency dependence of $\phi_d$ of $d$-orbital at ${\bf K}=(\pi,0)$ (filled) and ${\bf K}=(0,\pi)$ (unfilled).
The comparison with $1-\lambda_d(T)$ at $T/t_d=0.125$ in Fig.~\ref{ldds} reveals that the moderate retardation nature of the $d$-wave pairing interaction is essential. In other words, the slower decaying rate of $\phi_d(i\omega_n)$ is generically detrimental to the pairing, which is a common feature shared with single-band model. For instance, the anisotropic $V_x=1.0$ has slowest decay which is consistent with the largest $1-\lambda_d(T)$ so that weakest pairing.
Besides, the decaying rate shows an abrupt change at large anisotropic $V_x/t_d > 0.5$ for both antinodal ${\bf K}$ directions. This matches with the observation of $1-\lambda_d(T)$ in Fig.~\ref{ldds} and -ImG$(\bf{K}, \omega=0)$ in Fig.~\ref{Gkw}.

\begin{figure} [h]

\psfig{figure=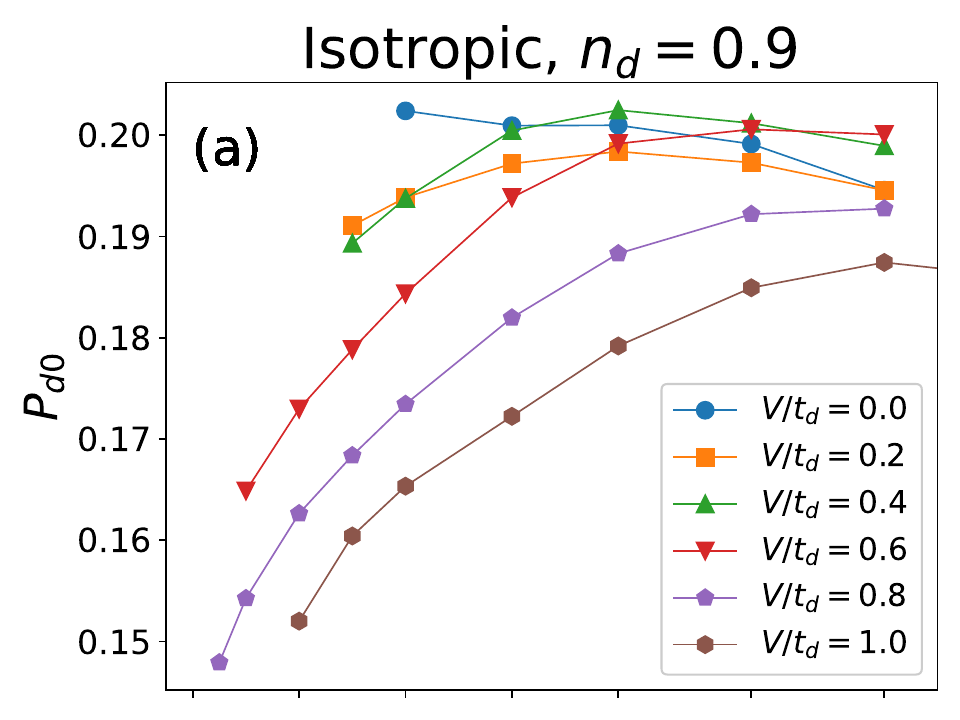,height=3cm,width=.23\textwidth, trim=12 1 10 0, clip} 
\psfig{figure=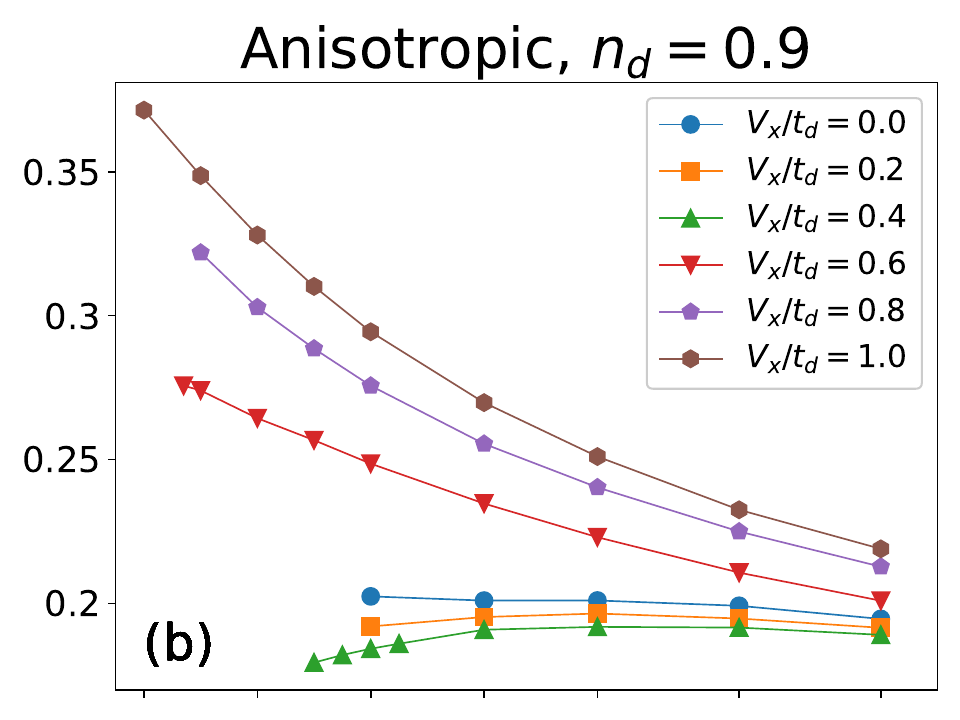,height=3cm,width=.23\textwidth, trim=8 1 -3 0, clip} 
\psfig{figure=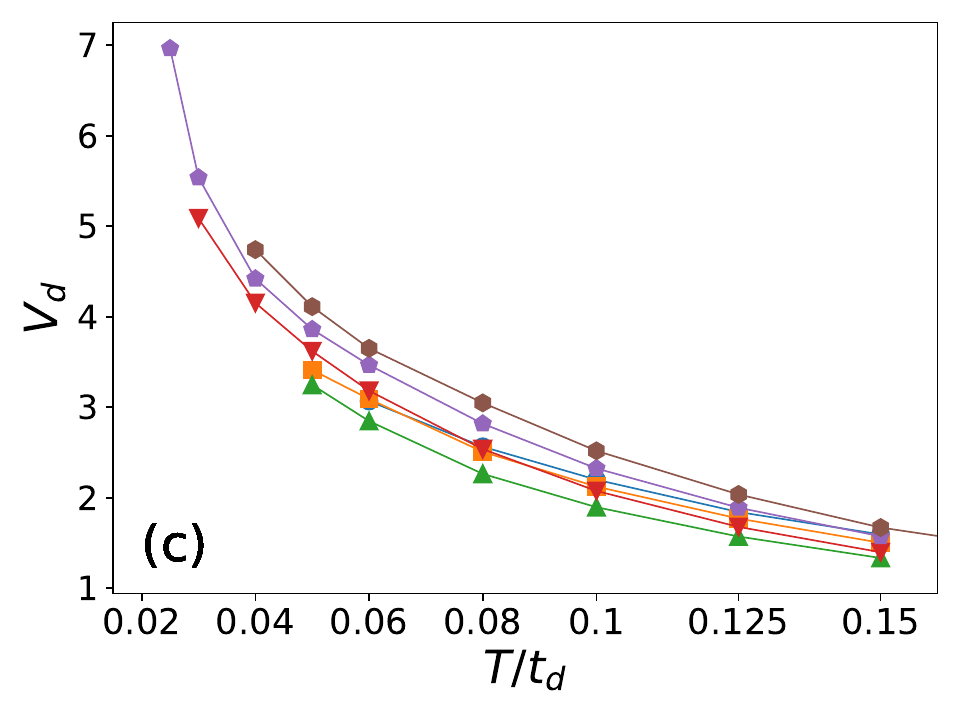,height=3cm,width=.23\textwidth, trim=-18 0 10 9, clip} 
\psfig{figure=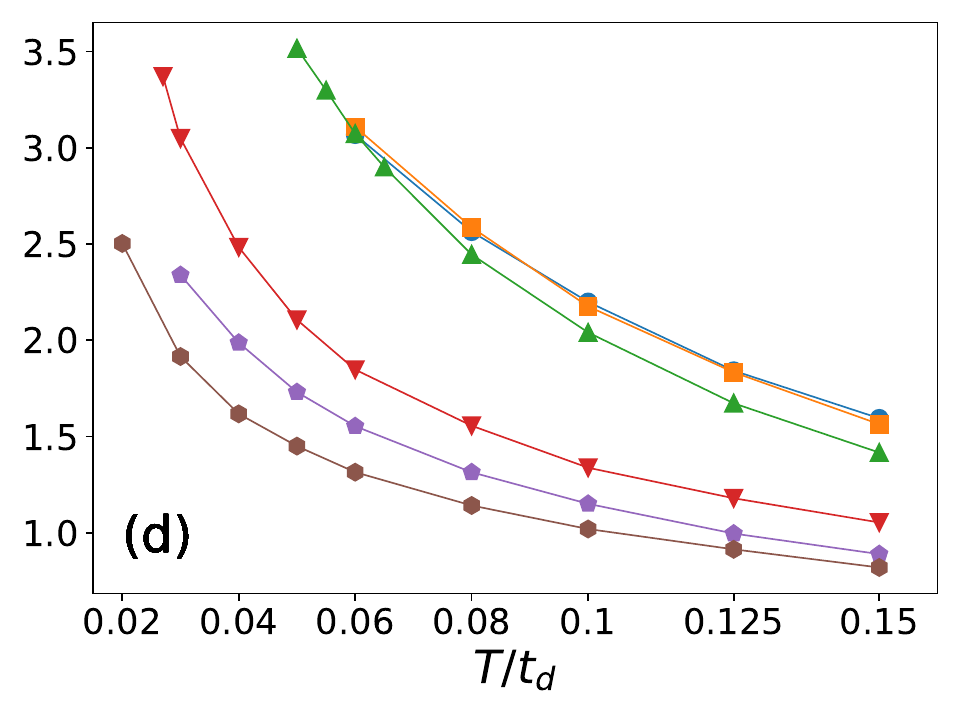,height=3cm,width=.23\textwidth, trim=-4 0 -3 9, clip} 

\psfig{figure=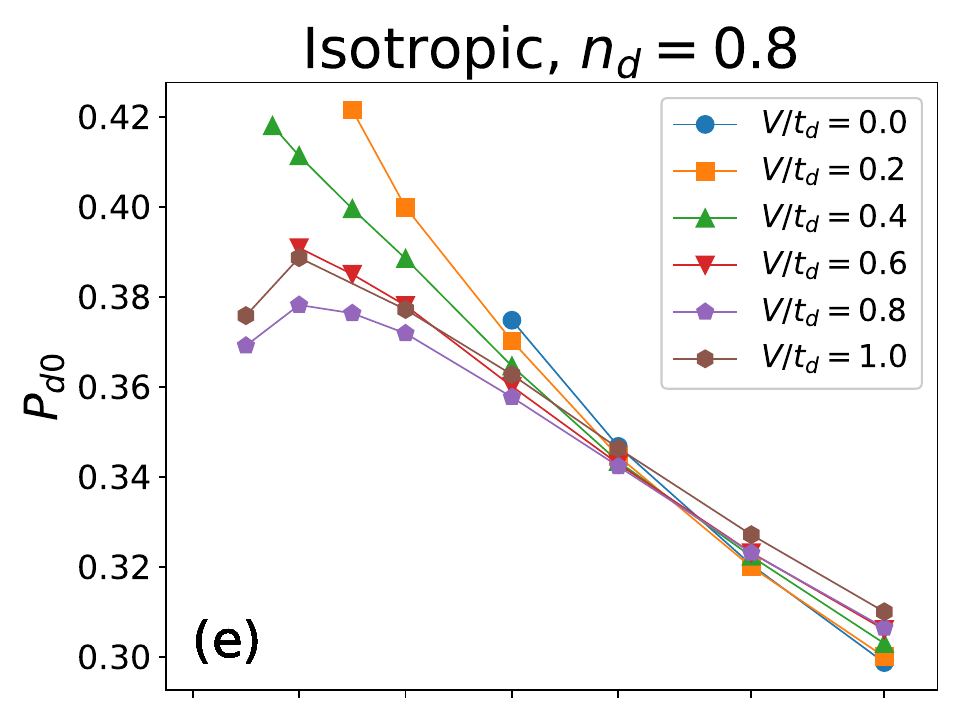,height=3cm,width=.23\textwidth, trim=12 0 10 0, clip} 
\psfig{figure=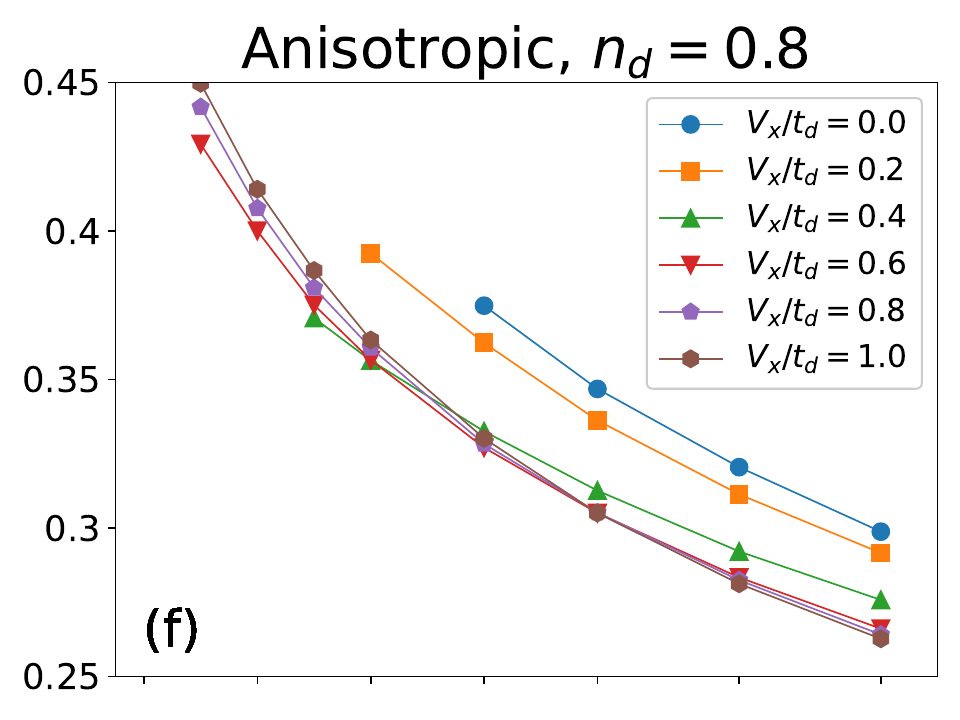,height=3cm,width=.23\textwidth, trim=8 7 -3 0, clip} 
\psfig{figure=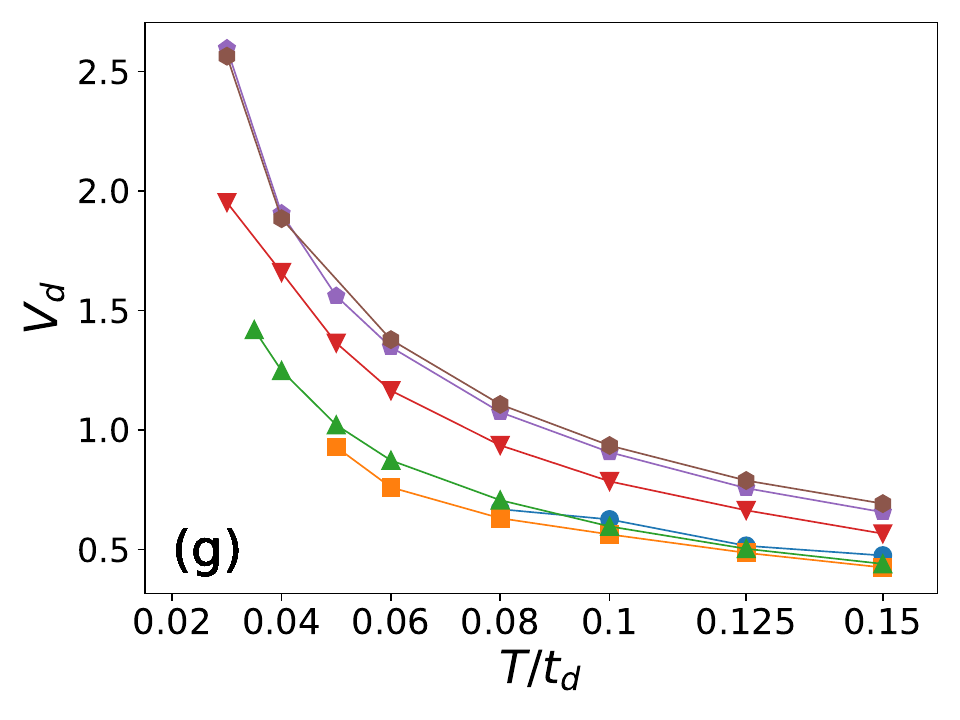,height=3.06cm,width=.23\textwidth, trim=0 0 10 9, clip} 
\psfig{figure=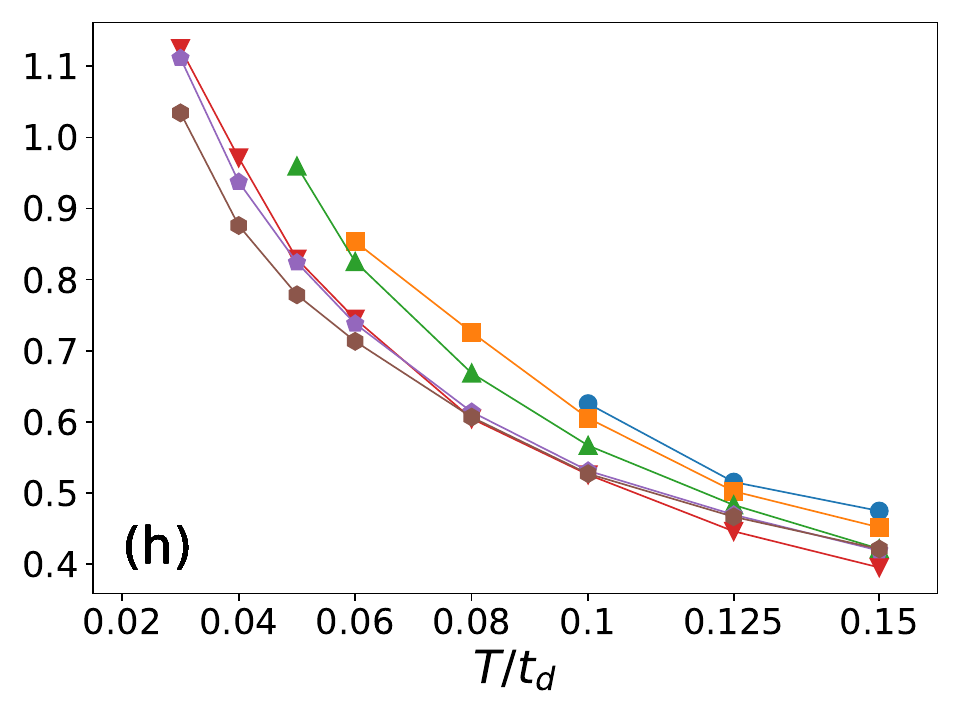,height=3.06cm,width=.23\textwidth, trim=-4 0 -3 9, clip} 

\caption{Temperature dependence of effective $d$-wave pairing interaction $V_d$ and bare pair-field susceptibility $P_{d0}$ for the $d$-$s$ model akin to Fig.~\ref{VdPd0hb}.}
\label{VdPd0ds}

\end{figure}

To gain deeper insight of the pairing interaction and intrinsic pairing instability, Fig.~\ref{VdPd0ds} shows that both isotropic $V$ and anisotropic $V_x$ induce opposite effects on $V_d$ and $P_{d0}$, which is distinct from the single band case in Fig.~\ref{VdPd0hb}. Moreover, the impact between isotropic and anisotropic hybridizations are opposite as well.
Hence, the compromise between $V_d$ and $P_{d0}$  leads to complicated dependence of the pairing instability upon $V(V_x)$.

At high $n_d=0.9$, although the isotropic $V$ (panel a) naturally suppresses $P_{d0}$ similar to the single band case, the anisotropic $V_x$ (panel b) counter-intuitively first suppresses but then promotes the intrinsic pairing instability with increasing $V_x$. This  might originate from the band structure reconstruction with $V_x/t_d > 0.5$, which is in fact also hinted in abrupt modification of $1-\lambda_d(T)$'s behavior in Fig.~\ref{ldds}(b) and Fig.~\ref{Gkw}(b). However, the variation of $P_{d0}$ will be compensated by the opposite trends of $V_d$, which should lead to roughly minor dependence of $T_c$ on $V(V_x)$ from Fig.~\ref{ldds}.

At low $n_d=0.8$, $V_d$ in panel (g) has relatively much larger increase compared to the moderate decrease of $P_{d0}$ in panel (e) with $V$. This leads to apparent boost of $1-\lambda_d(T)$ as observed in Fig.~\ref{ldds}(c). On the contrary, in the anisotropic case, both $P_{d0}$ and $V_d$ have only moderate variation with $V_x$ so that $1-\lambda_d(T)$ has minor dependence upon $V_x$ in Fig.~\ref{ldds}(d).
All these features indicate the complicated competition between anisotropy, hybridization, and doping range. The rapid change around $V_x/t_d=0.5$ might deserve further investigation in future.

{\em \textcolor{blue}{Summary and outlook:}}
In summary, by employing dynamic cluster quantum Monte Carlo calculations, we systematically explored the impact of broken rotational symmetry on the SC in the framework of two characteristic models with 2D Hubbard interactions representing the possible relevance to superconducting cuprates and nickelates respectively: (1) anisotropic nearest-neighbor hopping integrals and (2) anisotropic hybridization between an interacting $d$ orbital and a conducting (interstitial) $s$-band. Both models characterize the symmetry reduction from $C_4$ to $C_2$.
We discovered that both types of anisotropy destroy the $d$-wave pairing generically.  
However, in the anisotropic case, the high $n_d$ shows an abrupt change around $V_x/t_d \sim 0.5$ while the low $n_d$ hosts a weakly dependence of $d$-wave pairing tendency upon the anisotropy.
In addition, the evolution of the $d$-wave BSE eigenvector revealed that the moderate retardation is essential for the $d$-wave pairing in both models. In particular, the too slow decaying rate of $\phi_d(i\omega_n)$ with frequency is generically detrimental to the pairing. 

Our presented work matches with the experimental findings of the much lower superconducting $T_c$ of infinite-layer nickelates compared with the cuprates, which may be intrinsically connected to the anisotropy either from structural distortion or anisotropic hybridization with the interstitial orbitals. 
Furthermore, this exploration would contribute to the broader understanding of unconventional superconductors in the anisotropic environment.

{\em \textcolor{blue}{Acknowledgement:}}
We would like to thank Karsten Held for illuminating discussions.
H.~Z.~and M.~J.~acknowledge the support by National Natural Science Foundation of China (NSFC) Grant No.~12174278, startup fund from Soochow University, and Priority Academic Program Development (PAPD) of Jiangsu Higher Education Institutions. 
L.~S.~is thankful for the starting funds from Northwest University.
L.~S.~also acknowledges the financial support by projects P~32044 and I~5398 of the Austrian Science Funds (FWF).
Calculations have been done mainly on the computing facility in Soochow University, Super-computing clusters at Northwest University and Vienna Scientific Cluster (VSC).

\bibliographystyle{apsrev4-2}
\bibliography{main}

\end{document}